\documentstyle[preprint,aps]{revtex}
\begin{document}
\preprint{UCF-CM-96-001}
\title
{Ensemble density functional theory for inhomogeneous fractional quantum
Hall systems}
\author{O. Heinonen, M.I. Lubin, and M.D. Johnson}
\address{
Department of Physics, University of Central Florida, Orlando, FL 32816-2385
}
%\date{}
\maketitle
\begin{abstract}
The fractional quantum Hall effect (FQHE) occurs at certain magnetic
field strengths $B^*(n)$
in a two-dimensional electron gas of density $n$
at strong magnetic fields perpendicular to the plane of the electron gas.
At these magnetic fields strengths,
the system is incompressible, {\em i.e.,\/} there is a
finite cost in energy for creating charge density fluctuations in the
bulk, while the boundary of the electron gas has gapless modes of
density waves. The bulk energy gap arises because of the strong
electron-electron interactions.
While there are very good models for infinite homogeneous systems
and for the gapless excitations of the boundary of the electron gas,
computational methods to accurately model finite, inhomogeneous systems with
more then about ten electrons have not been available until very recently.
We will here review an ensemble density functional
approach to studying the ground state of large inhomogeneous spin polarized
FQHE systems.

\end{abstract}
\pagebreak

%\narrowtext
%\twocolumn
\section{Introduction}
The fractional quantum Hall effect (FQHE) is manifested in a two-dimensional
electron gas (2DEG) in a strong magnetic 
field oriented perpendicular to the plane of the 
electrons\cite{Girvin}. The effect was discovered as a transport anomaly. This
is still the `Hallmark' of the effect, even though there are now a host of other
phenomena associated with the effect which have been 
studied experimentally. In a transport
measurement 
it is noted that at certain strengths $B^*(n)$, which depend on the
density $n$ of the 2DEG, current can flow without any dissipation. 
That is, there
is no voltage drop along the flow of the current. At the same time, the
Hall voltage perpendicular to both the direction of the current and of the
magnetic field is observed to attain a quantized value for a small, but finite,
range of magnetic field or density, depending on which quantity is varied
in the experiment. 
The effect is understood to be the result of an excitation gap in the
spectrum of an infinite 2DEG at these magnetic fields, 
so that there is a finite cost in energy to be paid
for making density fluctuations in the system. This means that the 2DEG
is incompressible. In general, the magnetic field strengths $B^*(n)$ at
which the quantum Hall transport anomalies are observed 
are related to the density through the filling factor
$\nu=2\pi\ell_B^2 n$, with 
$\ell_B=\sqrt{\hbar c/(eB)}$ the magnetic length. 
The quantum Hall effect was first discovered\cite{Klitzing}
at integer filling factors. In this
integer quantum Hall effect, the energy gap is nothing but the kinetic
energy gap $\hbar\omega_c=\hbar eB/(m^*c)$.
Later, the fractional quantum Hall effect was discovered\cite{Stormer} 
at certain
rational filling factors of the form $\nu=p/q$, with $p$ and $q$ relative
primes, and $q$ odd. In the FQHE, the excitation gap is a consequence of
the strong electron-electron interactions.

Our understanding of the origin of the FQHE started with Laughlin's
seminal paper of 1983\cite{Laughlin}, which dealt with the simplest
fractions $\nu=1/m$, with $m$ an odd integer. At these values of $\nu$, there
are on the average $m$ magnetic flux quanta $\Phi_0=hc/e$ per electron. 
In that paper, Laughlin constructed a variational
wavefunction for spin-polarized systems in strong magnetic fields, strong enough
that
the splitting $\hbar\omega_c$ between the magnetic subbands, 
or Landau levels, can be taken to be infinite. The wavefunction can then
be constructed from single-particle states entirely within the lowest
Landau level. Laughlin wrote the variational wavefunction as
\begin{equation}
\Psi_m\propto\prod_{i,j}
\left(z_i-z_j\right)^m\exp\left[-\frac{1}{4}\sum_k|z_k|^2
\right]
\label{Laughlin:eq}
\end{equation}
where $m$ is an odd integer, and $z_j=x_j+iy_j$ is the coordinate of the $j$th
electron in complex notation. This wavefunction is an eigenstate of
angular momentum. Laughlin went on to demonstrate that the
system having the wavefunction Eq. (\ref{Laughlin:eq}) is an incompressible
liquid with $\nu=1/m$, $m$ odd, and 
with an energy gap to excitations, and that the elementary excitations are
fractionally charged quasi-holes or quasi-particles of charge $e^*=\pm e/m$.
The origin of the energy gap can be understood in the so-called
pseudo-potential representation of the electron-electron interactions
\cite{Haldane_pseudo}.
Here, the electron-electron interaction $V({\bf r}_i-{\bf r}_j)$ 
between electrons $i$ and $j$ is
decomposed into strengths $V_\ell$ in relative angular momentum channels 
$\ell=0,1,2,\ldots$,
of the two electrons. For any realistic interaction $V({\bf r}_i-{\bf r}_j)$,
it turns out that $V_0>V_1>V_2>\ldots$.
Consider the case $\nu=1/3$. In this case, 
the lowest angular momentum pseudo-potential that enters
into the $m=3$ Laughlin description
is $V_1$, the interaction energy of two electrons of unit 
relative angular momentum (in units of $\hbar$).
(Even relative angular momenta are not permissible for 
spin-polarized electron wavefunctions,
since they have to be anti-symmetric under 
interchange of electron coordinates.)
%(Zero relative angular momentum is not permissible for spin-polarized 
%systems due to the
%Pauli exclusion principle.) 
The Laughlin wavefunction is a very cleverly
constructed highly correlated state which
completely excludes unit relative angular momentum between
any two electrons, and is furthermore the only state which satisfies this
property at $\nu=1/3$.  Therefore, any excited state must contain some
electrons with unit relative angular momentum. The energy gap is due to
the cost of this, and hence is of order $V_1$. 
Note that because of the nature of the
correlations between the electrons, which are contained in the 
factors $(z_i-z_j)^m$, the Laughlin wavefunction cannot be expressed
as a single Slater determinant of single-particle states in the lowest
Landau level. Figure \ref{energy:fig} depicts the exchange-correlation
energy per particle for infinite, homogeneous FQHE systems {\em vs.}
filling factor. The cusps at 
filling factors $\nu=1/3,2/5,3/5$, and $\nu=2/3$ have been included to scale,
and these filling factors marked by vertical lines for clarity.
Note that these cusps are barely visible on this scale, yet they are
responsible for all the physics of the FQHE!

More modern theories of the FQHE are based on the so-called composite
fermions. This idea was originated by Jain,\cite{Jain}
who noted that the $m=3$
Laughlin wavefunction can be written as 
$\prod\left(z_i-z_j\right)^2\Psi_1$, where
$\Psi_1=\prod(z_i-z_j)\exp\left[-\frac{1}{4}\sum|z_k|^2\right]$ is the Slater
determinant wavefunction of a filled lowest Landau level. Because of 
constraints on the Hilbert space, $\Psi_1$ is an exact eigenstate
of an interacting electron system at $\nu=1$.
Although $\Psi_1$ need not in general
be the {\em ground state} of the interacting system, it is if 
the Landau level
splitting $\hbar\omega_c$ is much greater than the scale of the Coulomb
interaction, $e^2/(\epsilon_0\ell_B)$, and if the external
potential is sufficiently well behaved\cite{MacDonald_arg}, 
{\em e.g.,\/} it is caused by a
uniform positive background charge density, and we will assume that these
conditions are satisfied.
Multiplication by the Jastrow factor
$\prod(z_i-z_j)^2$  makes the total wavefunction $\Psi_3$ vanish as
the cube of the separation of two electrons when they approach each other,
rather than linearly. In the two-dimensional world of the FQHE, a zero
of the wavefunction at a fixed point $z_0$ produces a phase factor of $2\pi$ 
when an electron adiabatically encircles $z_0$. This is 
equivalent to adding a flux-tube containing a single flux quantum $\Phi_0$
at $z_0$. This is purely an Aharonov-Bohm phase -- no magnetic field is
added anywhere except at $z_0$, where the wavefunction vanishes. 
Therefore,
the FQHE at $m=3$ can be interpreted as a system of electrons at
$\nu=1$, but with two flux quanta added to the position of each electron.
These composite objects, electrons plus an integer number of flux quanta,
are called composite fermions. So one can say that the fractional
quantum Hall effect at $\nu=1/3$ is an integer quantum Hall effect ($\nu=1$)
of composite fermions. This generation of FQHE states from composite fermions
was subsequently generalized to all FQHE fractions. One way of studying 
FQHE systems of electrons theoretically is to perform a singular 
gauge transformation\cite{Chern}, 
which is a gauge transformation in which an odd
number of flux quanta is added to the position of each electron. The
transformed wavefunction is thus a wavefunction for composite fermions.
In a Lagrangian formulation, a term has to be added to the Lagrangian
to ensure that the flux tubes indeed are located
at the electrons. The resulting term in the Lagrangian is called the
Chern-Simons term, and is well known from earlier topological 
field theories.

\section{Finite systems}
We have outlined above how the
electron-electron interactions in an infinite, homogeneous system
produce the excitation gap. It is important to note that these gaps are only
for excitations in the {\em bulk} of the system. 
When a system is bounded there {\em must} be gapless excitations located
at the boundaries of the system. The following simple argument, due to
MacDonald\cite{MacDonald_arg}, 
demonstrates this point. Consider a finite system in which the
chemical potential $\mu$ lies in the bulk `charge gap', 
{\em i.e.,\/} we have to pay
the price of the energy gap to introduce particles to the bulk of the
system. Now imagine that the chemical potential 
is increased an infinitesimal amount 
$\delta\mu$, and consider the resulting change in the current density.
In the bulk, the current density cannot change since $\delta\mu$ is infinitesimal
and cannot overcome the energy gap in the bulk. It follows that if there
is a change in the current density as a response to $\delta\mu$, this must
be located at the edges of the system. Current conservation also requires that
if there is a resulting change in the current along the edge, this
change must be constant along the edge. We can relate the change in current
$\delta I$ to the change in orbital magnetization through
\begin{equation}
\delta I=\frac{c}{A}\delta M,\label{I-M:eq}
\end{equation}
where $A$ is the total area of the system. This relation is nothing but the
equation for the magnetic moment of a current loop. But we can write
$\delta M$ in terms of $\delta\mu$ using a Maxwell relation:
\begin{equation}
\delta M=\left.{\partial M\over\partial\mu}\right|_B\delta\mu=\left.{\partial N
\over\partial B}\right|_\mu\delta\mu,
\label{delta_M:eq}
\end{equation}
where $N$ is the total number of electrons. By combining Eqs. (\ref{I-M:eq})
and (\ref{delta_M:eq}) we arrive at
\begin{equation}
{\delta I\over\delta\mu}=c\left.{\partial n\over\partial B}\right|_\mu.
\end{equation}
Since $\partial n/ \partial B$ is non-zero, it follows that $\delta I/\delta\mu$
is non-zero, and we conclude that there must be gapless excitations of the
system, and that these excitations must localized to the edges. Since all
experimental systems are finite and inhomogeneous, the 
low-energy properties probed by experiments
must be determined by the gapless edge excitations. Advances in 
semiconductor nanofabrication technologies have lead to the possibility
of manufacturing systems which are extremely inhomogeneous, and in practice
dominated by edges. As an example, recent experiments
have even been performed on tiny quantum dots, with about 30 electrons
on them\cite{McEuen,Klein}. 
In order to accurately understand the experiments and inhomogeneous
FQHE systems in general, we must have a way of accurately calculating their
properties.
Certain aspects of inhomogeneous FQHE systems have been studied by different
techniques. For example, field theories can be constructed to study the
low-energy limit of the gapless edge excitations\cite{Wen}. Composite
fermion methods have been used in a Hartree approximation to study
finite FQHE systems\cite{Brey,Chklovskii2}. 
In this approach, the Chern-Simons term, arising from
the singular gauge transformation, and the electron-electron interaction
are treated in a self-consistent
Hartree approximation. The hope is then that the most important aspects of the
electron-electron correlations are included in this approximation.
Near $\nu=1$, at which the Slater determinant $\Psi_1$ 
is the exact ground state, it makes sense to use the Hartree-Fock approximation,
and the stability of a quantum dot at $\nu=1$ as a function of confining
potential has been studied in this approximation\cite{MacD_Yang,Chamon}. 
The edge structure has also been studied using semiclassical 
methods\cite{Beenakker,Chklovskii}, in 
which the electron-electron interaction is included at the Hartree level and
it is furthermore assumed that all potentials vary on a length scale much
larger than $\ell_B$. Beenakker\cite{Beenakker}, 
and Chklovskii, Shklovskii and Glazman\cite{Chklovskii} 
demonstrated that the edge of an
integer quantum Hall system, in which the correlation energies between the
electrons can be ignored, consists of a sequence of compressible and
incompressible strips. Imagine going from the bulk of an integer quantum
Hall effect, with $\nu$ filled Landau levels in the bulk (so that the
chemical potential lies above the energy of these $\nu$ Landau levels), 
toward an edge.
There is an external potential confining the system, and this potential
rises toward the edge, causing the Landau levels to bend upward near the edge.
The compressible strips occur where the
chemical potential crosses a Landau level. There are then both empty and
occupied 
single-particle states available, and the electron gas can screen the
external potential perfectly. Eventually this Landau level bends upward,
rising above the chemical potential. There are then
no more empty single particle states, and changing the electron density
would involve a cost of energy of the order of $\hbar\omega_c$, the
spacing between the Landau levels. Thus, an incompressible strip forms.
Here, the electrons cannot screen the external potential. Further out
on the edge, the
chemical potential crosses the next lower Landau level, and a new compressible
strip forms. Thus, whenever the chemical potential lies between two Landau
levels, the electron gas is incompressible, and the electron density is
constant, while the total potential varies. On the other hand, whenever the
chemical potential crosses a Landau level, the electron gas can screen perfectly
the external potential, and the electron density varies while the
total potential remains constant. The width of the incompressible
strips is then determined by the length over which the confining
potential varies an amount equal to the energy gap $\hbar\omega_c$.
The origin of the compressible and
incompressible strips are the energy gaps, which are the kinetic energy
gaps $\hbar \omega_c$ in the case of the integer quantum Hall effect. But
it is easy to heuristically generalize the 
argument to include the energy gaps causing
the FQHE\cite{Chklovskii}. 
The conclusion is then that there should be compressible
and incompressible strips, with the density of the incompressible strips
fixed at the density of an FQHE fraction. The width of each incompressible
strip is then fixed by the length over which the confining potential
varies an amount equal to the energy gap of the FQHE fraction corresponding to
the density of that strip.

Finite, inhomogeneous systems have also been studied by direct numerical
diagonalizations\cite{Yang_MacDonald_Johnson}.
At the present, numerical diagonalizations are limited to systems with
of the order of 10 electrons. 

It is highly desirable to have a 
computational approach which accurately includes electron-electron
correlations and can handle inhomogeneous systems  
with on the order of 
$10^2$--$10^3$ electrons.
One such approach which is in principle valid
for any interacting electron system is the 
density functional theory (DFT)\cite{Kohn_Vashista,Dreizler,Parr}. 
There have been some attempts to apply density functional theory to the
FQHE. Ferconi and Vignale\cite{Ferconi_CDFT} applied current density
functional theory\cite{Vignale_CDFT} to small, parabolically confined
quantum Hall systems and showed that the current density functional theory gave
good results for the ground state energy and spin polarization near $\nu=1$.
However, the energy gaps due to correlation effects were not included
in that calculation. Ferconi, Geller and Vignale\cite{Ferconi} 
also recently studied
FQHE systems within the spirit of the DFT using an extended 
Thomas-Fermi approximation at low, but non-zero, temperatures. 
In this, the kinetic energy was treated as
a local functional, as in the standard Thomas-Fermi approximation, while the
exchange-correlation energy was included 
in a local density approximation (LDA). 
This extended Thomas-Fermi approximation is valid in the limit
of very slowly varying confining potential. Ferconi, Geller and Vignale
focused on the incompressible and compressible strips at an edge of an
FQHE system, and obtained results in agreement with the predictions
by Chklovskii, Shklovskii, and Glazman\cite{Chklovskii}.

We have developed for the fractional quantum Hall effect 
an ensemble DFT scheme within the local density 
approximation, 
and have applied it to spin-polarized 
circularly
symmetric quantum dots\cite{Heinonen}. In our approach, the kinetic
energy is treated exactly, and the density represented by Kohn-Sham 
orbitals.
The results are in good agreement with results
obtained by semiclassical\cite{Beenakker,Chklovskii,Ferconi}, 
Hartree-Fock\cite{MacD_Yang,Chamon} (for cases where the
correlations do not play a major role), and exact diagonalization 
methods\cite{Johnson}.
Our calculations show that the exchange and
correlation effects of the FQHE are very well represented by the LDA and that
our approach provides a computational scheme to model large inhomogeneous
FQHE systems. We note that
there exist previous formal DFTs 
for strongly correlated systems, in particular for high-temperature
superconductors \cite{Gross}, and DFT calculations of high-T$_{\rm c}$
materials \cite{Pickett} and transition-metal oxides \cite{Svane}.
However, ours are, to the
best of our knowledge, the first practical LDA-DFT calculations of
a strongly correlated system in strong magnetic fields, and demonstrate the
usefulness of the LDA-DFT in studying large inhomogeneous FQHE systems.

\section{Ensemble density functional theory approach}
In typical DFT calculations of systems of $N_{\rm el}$ electrons, 
the standard Kohn-Sham (KS) scheme\cite{KohnSham} is implemented, 
in which the particle density $n({\bf r})$ is expressed in terms of
a Slater determinant of $N\geq N_{\rm el}$ 
KS orbitals, $\psi_{\alpha}({\bf r})$. These obey an effective
single-particle Schr\"odinger equation 
$H_{\rm eff}\psi_\alpha=\epsilon_\alpha\psi_\alpha$, which is solved 
self-consistently by occupying the $N_{\rm el}$ KS orbitals with the lowest
eigenvalues $\epsilon_\alpha$,
and iterating. 
This scheme works well in practice
for systems 
for which the true 
electron density can be represented by a single Slater
determinant of single-particle wavefunctions. 
However, when the KS orbitals are degenerate at the Fermi energy
(which we identify with the largest $\epsilon_\alpha$ of the occupied
orbitals)
there is an
ambiguity in how to occupy these degenerate orbitals.
There exists an extension of DFT which is formally able to deal
with this situation. This extension is called ensemble DFT\cite{Dreizler,Parr}, 
and in it,
the density of the system is represented by an ensemble of Slater
determinants of KS orbitals. However, while it can be  shown using 
ensemble DFT 
that such a representation of the density is rigorous, it cannot be shown
{\em how} the degenerate KS orbitals at the Fermi energy should be occupied,
{\em i.e.,\/} there has not been available a practical computational scheme
for ensemble density functional theory.  

We argued above that the Laughlin wavefunction cannot be represented
as a Slater determinant of single-particle wavefunctions. Therefore, one
may suspect that the density of a general FQHE system cannot be represented
by a single Slater determinant. We will now argue that this is indeed the case.
Consider a FQHE system in
the $xy$-plane with the magnetic field along the $\hat z$-axis. 
A circularly symmetric external potential
$V_{\rm ext}({\bf r})=V_{\rm ext}(r)$ 
(due, {\em e.g.,\/} to a uniform positive background charge density) 
confines the systems such that the density
is fixed with a local filling of $\nu=1/3$ up to an edge at $r_0$ ($r_0\gg\ell_B$)
where the density falls to zero within a distance of order $\ell_B$.
That such
systems exist is well demonstrated by the excellent agreement
between the Laughlin wavefunction and experiments, and by many
numerical
calculations\cite{Haldane,Johnson}.
Due to the circular symmetry
we can label 
single-particle orbitals by 
angular momentum $m$, and by a Landau level index $n\geq0$. 
The orbitals $\psi_{m,n}({\bf r})$ are centered
on circles of radii $r_m\approx\sqrt{2m}\ell_B$ with Gaussian fall-offs
for $r\ll r_m$ and $r\gg r_m$. 
The single-particle orbitals with $n=0$ are then in the 
bulk all degenerate, and the
degeneracy is not lifted by electron-electron interactions since the system is
homogeneous in the bulk. Only the orbital $\psi_{0,0}$ is non-zero at
the origin -- all others vanish at $r=0$.
In order to obtain a constant density at $\nu=1/3$ even at the 
center of the system,
all single-particle orbitals  in the bulk with $n=0$ must 
have occupancies $1/3$. If the Fermi energy lay above the energies of the
bulk orbitals, they would all be filled and one would have $\nu=1$.
Therefore, to get occupancies $1/3$ the Fermi energy must lie at 
the degenerate energy $\epsilon_{m0}$
of these orbitals.
Thus, 
in applying DFT to the FQHE we can expect a huge degeneracy of KS orbitals
at the Fermi energy, 
and they must
all have fractional occupancies. 
Consequently, the particle density cannot be expressed
in terms of a single Slater determinant. Instead,
the density has to be constructed from an ensemble of Slater determinants, 
{\em i.e.,\/} the orbitals at the Fermi energy are assigned fractional
occupation numbers, just as
the Laughlin wavefunction for $\nu=1/3$ is not a single Slater determinant,
but a highly correlated state with average occupancies of 1/3 of
single-particle states. Therefore, one might expect from the outset that
one has to use ensemble density functional theory -- the standard Kohn-Sham
scheme may not converge.

Although ensemble DFT has been developed formally, there are in practice
few examples of applications and calculations using ensemble DFT
for ground state calculations. A significant aspect of our
work is that we have developed an ensemble scheme which is practical 
and useful
for the study of the FQHE.
In ensemble DFT, any physical density
$n({\bf r})$ can be represented by
$
n({\bf r})=\sum_{mn}f_{mn}|\psi_{mn}({\bf r})|^2,
$
where $f_{mn}$ are occupation numbers satisfying
$0\leq f_{mn}\leq1$, and the orbitals $\psi_{mn}$
satisfy the equation
\begin{equation}
\left\{
\frac{1}{2m^*}\left[{\bf p}+\frac{e}{c}{\bf A}({\bf r})\right]^2
+V_{\rm ext}({\bf r})+V_{\rm H}({\bf r})
+V_{\rm xc}({\bf r},{\bf B})\right\}\psi_{m,n}({\bf r})=\epsilon_{mn}
\psi_{mn}({\bf r}),\label{HK}
\end{equation}
where $\nabla\times{\bf A}({\bf r})={\bf B}({\bf r})$.
In equation (\ref{HK}),  
$V_{\rm H}({\bf r})$ is the 
Hartree interaction of the 2D electrons, and, as usual, 
$V_{\rm xc}({\bf r},{\bf B})$ is the exchange-correlation potential, defined as
a functional derivative of the exchange-correlation energy 
$E_{\rm xc}[n({\bf r}),{\bf B}]$ 
of the system with respect to density:
$
V_{\rm xc}({\bf r},{\bf B})=\left.{\delta E_{\rm xc}[n({\bf r}),{\bf B}]
\over\delta n({\bf r})}\right|_{\bf B}.
$
(We will hereafter not explicitly indicate the parametric dependence
of $V_{\rm xc}$ and
$E_{\rm xc}$ on $\bf B$.) For the case of the FQHE, we know that the
exchange-correlation potential will be crucial, as it contains all the effects
of the electron correlations which cause the FQHE in the first place, and
a major part of the DFT application is to come up with an accurate model
of $E_{\rm xc}$ and so of $V_{\rm xc}$. Leaving this question aside for a moment,
and assuming that we have succeeded in doing so,  
the practical question is then how to determine
the KS orbitals and their occupancies in the presence of degeneracies. We
devised an empirical scheme, which means that after a lot of trial and error
we made some educated guesses that work. Our scheme produces a set of
occupancies for the KS orbitals which satisfy some minimum requirements, namely
(a) the scheme converges
to physical densities (to the best of our knowledge) for FQHE systems, 
(b) it reproduces finite temperature DFT distributions at finite
temperatures, and (c) it reproduces the standard Kohn-Sham scheme for
systems whose densities can be represented by a single Slater determinant.

In our
scheme, we start with input occupancies and single-particle orbitals and
iterate the system $N_{\rm eq}$ times using the KS scheme. 
The number $N_{\rm eq}$ is chosen large enough (about 20--30 
in practical calculations)
that the density is close to the final density after $N_{\rm eq}$ iterations. 
If the density of the system could be represented by a single
Slater determinant of the KS orbitals, we would now essentially
be done. However, in this system there are now in general many degenerate
or near-degenerate orbitals at the Fermi energy. After each iteration, the
Kohn-Sham scheme chooses to occupy the $N_{\rm el}$ orbitals with the lowest
eigenvalues, corresponding to making a distinct Slater determinant of these
orbitals. But there will be small fluctuations in the density between each
iteration, which cause a different subset of these (near) degenerate
orbitals to be occupied after each iteration.
This corresponds to constructing
different Slater determinants after each iteration, and the occupation numbers
$f_{mn}$ of these orbital are zero or unity more or less at random after
each iteration. This means that the computations will never converge.
However, the {\em average} occupancies, {\em i.e.,\/} the
occupancies averaged over many iterations, become well defined and approach
a definite value, {\em e.g.,\/} 1/3 for orbitals localized 
in a region where the local filling factor is close to $\nu=1/3$.
Therefore, we use these average occupancies to construct an ensemble by
accumulating running average occupancies $\langle f_{mn}\rangle$
after the initial $N_{\rm eq}$ iterations 
\begin{equation}
\langle f_{mn}\rangle=
{1\over(N_{\rm it}-N_{\rm eq})}\sum_{i=N_{\rm eq}+1}^{N_{\rm it}}f_{mn,i},
\label{average_occ:eq}
\end{equation}
where $f_{mn,i}$ is the occupation number (0 or 1) of orbital $\psi_{mn}$ 
after the $i$th iteration,
and use these to calculate densities. 
Thus, our algorithm essentially
picks a different (near) degenerate Slater determinant after each 
iteration, and these
determinants are all weighted equally in the ensemble.
It is clear that this scheme reduces to the KS scheme for which the
density can be represented by a single Slater determinant of KS orbitals
(for which the KS scheme picks only the one Slater determinant which gives
the ground state density)
for $N_{\rm eq}$ large enough. We have numerically 
verified that a finite-temperature version of our scheme converges
to a thermal ensemble at finite temperatures down to
temperatures of the order of $10^{-3}\hbar\omega_c/k_B$. We
have also performed some Monte Carlo simulations about the
ensemble obtained by our scheme. In these simulations, we used a Metropolis
algorithm to randomly change the occupation numbers about our converged
solution, keeping the chemical potential fixed. The free energy of the
new set of occupation numbers was calculated self-consistently. 
If the free energy decreased,
this set was kept, and if the free energy increased, the set was kept
if a random number was smaller than $\exp\left[-\Delta F/k_BT^*\right]$, where
$\Delta F$ is the change in free energy, and $T^*$ a fictitious temperature. 
The results were that to within numerical
accuracy our ensembl DFT scheme gives the lowest free energy.
As a condition for convergence, we typically demanded that the difference
between the input and output ensemble densities, $n_{\rm in}(r)$ and
$n_{\rm out}(r)$, of one iteration should satisfy
\begin{equation}
{1\over N_{\rm el}}\int_0^\infty\,\left|n_{\rm in}(r)-n_{\rm out}(r)\right|
r\,dr<10^{-3}.
\end{equation}

Practical density functional theory calculations hinge on the availability
of good approximations for the exchange-correlation potential $V_{\rm xc}$,
which enters in the effective Schr\"odinger equation for the KS orbitals.
The simplest, and probably the most commonly used, approximation is the
local density approximation (LDA).
In this approximation, the exchange-correlation
energy is assumed to be a {\em local function} of density, so that
the total exchange-correlation energy consists of contributions from the
local density of the system. Thus, in this approximation one writes
$
E_{\rm xc}/N=\int d{\bf r}\epsilon_{\rm xc}(\nu)n({\bf r}),
$
where $\epsilon_{\rm xc}(\nu)$ is the exchange-correlation energy per
particle in a {\em homogeneous} system of constant density 
$n=\nu/(2\pi\ell_B^2)$ and 
filling factor $\nu$. In other words, in the LDA one assumes that the
system is locally homogeneous, {\em i.e.,\/} the system can locally
be approximated to have the energy per particle of an infinite, homogeneous
system of the local density. This approximation obviously makes sense
if the density of the system varies on a very long length scale, while
it could be questionable for systems in which the density varies on
some microscopic length scale. However,  
experience has shown that the LDA often works surprisingly well,
even
for systems in which the electron density is strongly 
inhomogeneous\cite{Kohn_Vashista}. In fact, the first application of 
LDA-DFT was to calculate the work function of simple metals\cite{Lang}, so these
were systems which were terminated with densities varying on the scale of
a Bohr radius! Nevertheless, the LDA-DFT gave quite good results, vastly
superior to those of the Hartree- or Hartree-Fock approximation.
In the
FQHE, the length scale of exchange-correlation interactions 
and density fluctuations is
given by the magnetic length $\ell_B$ due to the Gaussian fall-off of
any single-particle basis in which the interacting Hamiltonian
is expanded. The densities are relatively smooth on this length scale,
which gives us additional hope that the LDA will work well for the FQHE, too.
In addition, the cusps in the exchange-correlation energy will suppress density
fluctuations, so in this sense one can actually expect the basic physics
of the FQHE to make the LDA a good approximation.

In conventional LDA-DFT calculation, the exchange-correlation energy
$\epsilon_{\rm xc}$ is obtained by interpolating between the exchange-correlation
energies per particle 
of systems with vanishing and infinite densities, respectively,
for which exact results are known. Analogously, following Rasolt and Perrot
\cite{Rasolt}, we obtain our
exchange-correlation energy by interpolating between two limits for which
the result is known very accurately. In our case, the two
limits are $B\to\infty$, and $B\to0$, respectively, and we stitch them
together using a Pad\'e approximant\cite{Rasolt}.
Thus, we write for the exchange-correlation energy per
particle of
a uniform electron gas in a constant magnetic field
\begin{equation}
\epsilon_{\rm xc}(\nu)={\epsilon_{\rm xc}^{\rm FQHE}(\nu)+
\nu^4\epsilon_{\rm xc}^{\rm TC}(n(\nu))\over 1+\nu^4}.
\end{equation}
Here, $\epsilon_{\rm xc}^{\rm TC}$ is the zero-magnetic field result 
for a 2DEG obtained by Tanatar and Ceperley
\cite{Tanatar}. 
The term $\epsilon_{\rm xc}^{\rm FQHE}(\nu)$ is the $B\to\infty$ limit, which
is the exchange-correlation energy of the FQHE in a system for which
only single-particle states in the lowest Landau level are occupied.
This contribution consists of two terms. 
The first one is a smooth interpolation
formula $\epsilon_{\rm xc}^{\rm LWM}(\nu)$ 
due to Levesque, Weiss and MacDonald \cite{Levesque}
between ground state 
energies at some rational fillings. The
second one, $\epsilon_{\rm xc}^{\rm C}(\nu)$, 
is all-important for the study of the FQHE. This term contains the
cusps in the ground state energy which cause the FQHE. Here we have used
a simple model which captures the essential physics. We model
$\epsilon_{\rm xc}^{\rm C}(\nu)$ by constructing it to be zero at
values of $\nu=p/q$ which display the FQHE. Near $\nu=p/q$, 
$\epsilon_{\rm xc}^{\rm C}(\nu)$ is linear and
has at $\nu=p/q$ a discontinuity in the
slope related to the chemical potential gap 
$\Delta \mu=q(|\Delta_p|+|\Delta_h|)$. Here $\Delta_{p,h}$ are the 
quasiparticle (hole) creation energies  which
can be obtained from the literature \cite{Morf_Halperin,Morf_Ambrumenil}
at fractions $\nu=p/q$.
Farther
away from $\nu=p/q$, $\epsilon_{\rm xc}^{\rm C}(\nu)$ decays to zero.
Finally, in
the LDA $V_{\rm xc}(r)$ is obtained from $\epsilon_{\rm xc}(\nu)$ as
$
V_{\rm xc}(r)=\left.{\partial \left[\nu\epsilon_{\rm xc}(\nu)\right]
\over\partial\nu}\right|_{\nu=\nu(r)}
$
at constant $B$. In our calculations, we restrict ourselves to include
only the cusps at $\nu=1/3,2/5,3/5$ and $\nu=2/3$, which are the
strongest fractions. These are some the fractions 
of the form $
\nu={p\over(2p\pm1)}$ generated by the  
so-called $V_1$-model, in which only the pseudo-potential $V_1$ is included.

A technical difficulty arises in the LDA: the 
discontinuities in $V_{\rm xc}(r)$ in the LDA give rise to a 
numerical instability. The reason
is that an arbitrarily small fluctuation in charge density close to an FQHE
fraction gives rise to a finite change in energy. Imagine that
the local filling factor $\nu(r)$ 
in some neighborhood of a point $r$ is very close to,
but less than, say, 1/3 after one iteration. In this neighborhood, the 
local exchange-correlation potential will then form a potential well
with
sharp barriers at the points around $r$ where $\nu(r)=1/3$. During
the next iteration, charge will then be poured into this well. As a
result, the local filling factor will after this iteration exceed 
1/3, and in this neighborhood $V_{\rm xc}$ now forms a potential barrier of
finite height.
So in the next iteration, charge is removed from this
neighborhood, and so on. We can see that this leads to serious
convergence problems. To overcome this,
we made the compressibility of the system finite, but very small, 
corresponding to a finite, but very large, curvature instead of a point-like 
cusp in $\epsilon_{\rm xc}$ at the FQHE fractions. In other words,
instead of having a step-like discontinuity $\Delta\mu$ in the chemical potential,
it rises smoothly an amount $\Delta\mu$ over an interval $\gamma$ in the
filling factor. What we found worked very
well in practice was to have the discontinuity in chemical potential occur
over an interval of filling factor $\gamma$ of magnitude 
$10^{-3}$. This corresponds
to a sound velocity of about $10^6$ m/s in the electron gas, which is
three orders of magnitude larger than the Fermi velocity of a 2D electron
gas at densities typical for the FQHE. In general, the finite compressibility
does not lead to any spurious physical effects so long as the energy of
density fluctuations on a size of the order of the systems size is larger
than any other relevant energy in the problem. The only noticeable
effect is that incompressible plateaus, at which the density would be
perfectly constant were the compressibility zero, will have density
fluctuations on a scale of $\gamma$. 
Figure \ref{V_xc} depicts  $V_{\rm xc}$ used in our calculations 
as a function of
filling factor.

\section{Applications to quantum dots}
We have self-consistently solved the KS equations Eqs. (\ref{HK})
for a spin-polarized quantum dot in a parabolic external potential, 
$V_{\rm ext}(r)=\frac{1}{2}m^*\Omega^2r^2$, by expanding the KS
orbitals $\psi_{mn}({\bf r})=e^{im\phi}\varphi_{mn}(r)$ in the eigenstates
of $H_0=\frac{1}{2m^*}\left({\bf p}+\frac{e}{c}{\bf A}({\bf r})\right)^2$.
We use the cylindrical gauge, ${\bf A}({\bf r})=\frac{1}{2}Br\hat\phi$, 
and include the four lowest Landau levels ($n=0,\ldots,3$). We
chose the static dielectric constant $\epsilon_0=13.6$, appropriate for
GaAs, and a confining potential of strength\cite{McEuen}
$\hbar\Omega=1.6$ meV.

The use of our LDA-DFT scheme is illustrated by a study of the
edge reconstruction of the quantum
dot as a function of magnetic field strength. As is known from
Hartree-Fock and exact 
diagonalizations\cite{MacD_Yang,Chamon,Johnson,Brey,Chklovskii2}, 
for strong confinement the 
quantum dot forms a maximum density droplet 
in which the density is uniform
at $\nu=1$ in the interior, and falls off rapidly to zero at 
$r\approx\sqrt{2N}\ell_B=r_0$. 
As the magnetic field strength increases,
a ``lump'' of density breaks off, leaving a ``hole'' or deficit at about
$r=r_0$. This effect is due to the short-ranged attractive
exchange interaction:
it is energetically favorable to have a lump of density break off so that the
system can take advantage of the exchange energy in the lump. As $B$ is
further increased, the correlations will cause
incompressible strips with densities $\nu=p/q$ to appear
\cite{Beenakker,Chklovskii,Gelfand,Ferconi} on the edges, 
and incompressible droplets to form in the bulk
at densities $\nu=p/q$. 
Figure \ref{reconstr} depicts various stages
of edge 
reconstruction obtained by us as the magnetic field strength is increased.
The value of 
$B$
for which the exchange lump appears compares very well with the value
found by De Chamon and Wen\cite{Chamon} 
in Hartree-Fock and numerical diagonalizations.
At higher fields still, incompressible strips appear
at the edges, and incompressible droplets are formed in the bulk.
In figure \ref{filling_B=4.45:fig} we show occupancies for the
KS orbitals for a finite-temperature calculation with
$N_{\rm el}=40$, $B=4.45$ T, and $T=0.003$ $e^2/(\epsilon_0\ell_Bk_B)$.
The diamonds depict the converged ensemble occupancies for the KS
single-particle states using our
ensemble scheme. At this finite temperature, we calculated the 
thermal occupancies of the KS orbitals after 
each iteration, and  these `instantaneous'
thermal occupancies were then averaged using our ensemble scheme. The 
temperature was sufficiently low that the `instantaneous' thermal
occupancies were essentially 0 or 1 before convergence. The continuous
curve shows the `instantaneous' thermal occupancies obtained after a particular
iteration after convergence has been achieved. This figures then clearly
shows that our ensemble occupancies (in this case at a low, but nonzero,
temperature) converged to the thermal occupancies. Note that this
particular temperature is so low that no standard finite-temperature scheme
could be used to achieve convergence.

Figure \ref{mu:fig} depicts the eigenvalues of the KS orbitals for
$N_{\rm el}=40$, and $B=4.45$ T. The dashed line indicates the chemical
potential of the system. This figure then shows that all KS orbitals in the
bulk
are in fact degenerate. It may at first seem paradoxical that the eigenvalues
are degenerate on an incompressible strip, since, according to the picture
by Chklovskii, Shklovskii, and Glazman\cite{Chklovskii}, 
on such a strip the density is constant, while
the total potential varies (since the electrons cannot screen the 
external potential). If the total potential varies, then ought not the
the eigenvalues of the KS orbitals localized on that strip vary, too, 
since these then in general are subjected to different potential energies?
The problem with this argument as applied to DFT is that it ignores the
effect of the exchange-correlation potential. As the external and
Hartree potentials vary across the strip, the exchange-correlation potential
varies across its discontinuity so as to completely screen out the
external and Hartree potentials. The discontinuity in $V_{\rm xc}$ does
{\em not} mean that this potential is fixed at the lower limit of its
discontinuity while the density is fixed at an incompressible strip. What it
does mean, is that $V_{\rm xc}$ is free to achieve any value across its
discontinuity so as to completely screen out the external and Hartree
potentials. In this way, it is perhaps better to think of incompressibility
as the limit of a finite compressibility approaching zero. A strip
can then remain incompressible with constant density so long as $V_{\rm xc}$
can screen the external and Hartree potentials, so the width of the
incompressible strip is given by the distance over which the external plus
Hartree potentials varies an amount given by the energy gap associated with
the density at that strip. 
Also, all bulk KS states are degenerate at the chemical potential. When a
single particle is added, the chemical potential simply increases a small
amount, and all KS orbitals are again degenerate at the chemical potential.
We also would like to
emphasize that incompressible regions that appear in these calculations
are not due to the presence of a
uniform positive background density which tends to fix the bulk density at
the value of the background density.

There is also another edge effect caused by
correlations. For particular, stiff confining potentials, so-called
composite edges\cite{MacDonald,Johnson} 
can appear. These can be thought of as particle-hole
conjugates of uniform incompressible droplets. 
Consider a droplet with a bulk density corresponding to $\nu=1/3$, 
falling off to zero at the
edge. A incompressible droplet with a bulk density of $\nu=2/3$ is 
obtained by particle-hole conjugation.
However, at the edge, the density will
first {\em rise} to $\nu=1$ (since the density of the $\nu=1/3$ droplet
drops to zero), and then eventually drop to zero. Note that this
argument is based on particle-hole conjugation, which is 
an exact symmetry of the lowest Landau level\cite{symmetry}, and it is unclear
if composite edges exist in real systems, which do not strictly obey
particle-hole symmetry.

Figure \ref{composite:fig} depicts 
the particle density (inset) for a system where the confining potential is
supplied by a uniform positive background charge density $n_+=2/(6\pi\ell_B^2)$ 
(so that the corresponding filling factor is $\nu_+=2/3$)
for $r<r_0$, and falling linearly to zero over a distance $a$ for $a>r_0$, where
$r_0$ is fixed by charge neutrality. Thus, the parameter $a$ is a 
convenient parameter with which one can control the stiffness of the
confining potential\cite{Chamon}.
From this figure, we see that for $a=0$,
the system forms a composite edge,  even though our system does
not obey particle-hole symmetry. We therefore conclude that such
edges can exist in real systems. As $a$ is increased, the edge undergoes
an instability and reconstruction, and eventually forms incompressible
strips.

\section{Conclusion and summary}
In conclusion, we have showed that ensemble density functional theory can
be applied to the FQHE. This opens the door to doing realistic calculations
for large systems. We believe that our results are also significant in that
they are the first LDA-DFT calculations of a strongly correlated system
in a strong magnetic field, and they are (to the best of our knowledge)
the first practical ensemble DFT calculations. There are, however,  
still many issues that need to be resolved, and new directions to go.
For example, our calculations were of a spin-polarized system. 
As is well known\cite{MacDonald_arg},
the spin degree of freedom is very important, even for magnetic fields
of the order of 10 T. The reason is that the effective $g$-factor in
GaAs is very small, so small that the ratio between the Zeeman splitting
and the cyclotron energy is about 2\%. This leads to the possibility
of FQHE ground states which are not spin-polarized. It also gives rise
to so-called charge--spin--texture excitations near $\nu=1/m$, $m=1,3,\ldots$. 
These excitations are lower in energy than a simple singlet particle-hole
pair, and believed to be responsible for the observed\cite{Barrett} 
rapid destruction
of the spin polarization near $\nu=1$. Correctly including the spin
degree of freedom to account for this involves a DFT for Heisenberg spins.
We are presently, together with J. Kinaret (Chalmers University of 
Technology) developing such a theory. 

The authors would like to thank M. Ferconi, M. Geller and G. Vignale for helpful
discussions and for sharing their results prior to publications, and 
K. Burke and E.K.U Gross for useful comments about the DFT. O.H. would like to 
thank Chalmers Institute of Technology, where part of the numerical work was done.
This work was supported by
the NSF through grant DMR93-01433.

\begin{figure}
\caption{ The ground state energy $\epsilon_{\rm xc}$ per particle
of an infinite, homogeneous, spin-polarized FQHE system is depicted 
as a function of
filling factor. The cusps at $\nu=1/3,2/5,3/5$, and $\nu=2/3$ are included
(these filling factors are indicated by vertical lines for clarity).}
\label{energy:fig}
\end{figure}
\begin{figure}
\caption{Exchange-correlation potential $V_{\rm xc}$ as function of
filling factor in units
of $e^2/(\epsilon_0\ell_B)$ for $0\leq\nu\leq1$.
The increase in $V_{\rm xc}$
at an FQHE filling occurs over a range of filling factor of 0.004.}
\label{V_xc}
\end{figure}
\begin{figure}
\caption{Edge reconstruction of a quantum dot as the magnetic field
strength is increased. 
Plotted here is the local filling factor 
$\nu(r)$ for a parabolic quantum dot with
$\hbar\Omega=1.6$ meV and
40 electrons. 
For magnetic field strengths $B< 2.5$ T the dot forms a maximum density
droplet, and for $B\approx3.0$ T, an exchange hole is formed. For stronger
magnetic fields, incompressible regions form, separated by compressible strips.}
\label{reconstr}
\end{figure}
\begin{figure}
\caption{Ensemble (diamonds) and `instantaneous' (solid line) 
thermal occupancies for 
$N_{\rm el}=40$, $B=4.45$T, and $T=0.003$ $e^2/(\epsilon_0\ell_Bk_B)$ after
convergence.}
\label{filling_B=4.45:fig}
\end{figure}
\begin{figure}
\caption{Eigenvalues of the lowest-Landau level 
Kohn-Sham orbitals for $N_{\rm el}=40$ and
$B=4.45$ T as a function of angular momentum quantum number. The chemical
potential is indicated by the dashed line.}
\label{mu:fig} 
\end{figure}
\begin{figure}
\caption{Local filling factor $\nu(r)$ as a function of $r$ 
(in units of $\ell_B$)
for 
a system of 45 electrons in a magnetic field of $B=5.0$ T. 
The confining potential is due to
a positive background charge density at $\nu_+=2/3$ in the bulk, and falling
linearly to zero within a distance $a$ near the edge.
For a stiff edge ($a=0$), the system forms a composite edge with $\nu(r)$
rising towards unity near the edge. As $a$ increases, the edge becomes softer
and undergoes a reconstruction to a sequence of incompressible strips.}
\label{composite:fig}
\end{figure}

\begin{references}
\bibitem{symmetry} Note, however, that for a finite system 
the Hilbert space for the holes
has to be restricted. See Ref. \onlinecite{Johnson}.
\end{references}
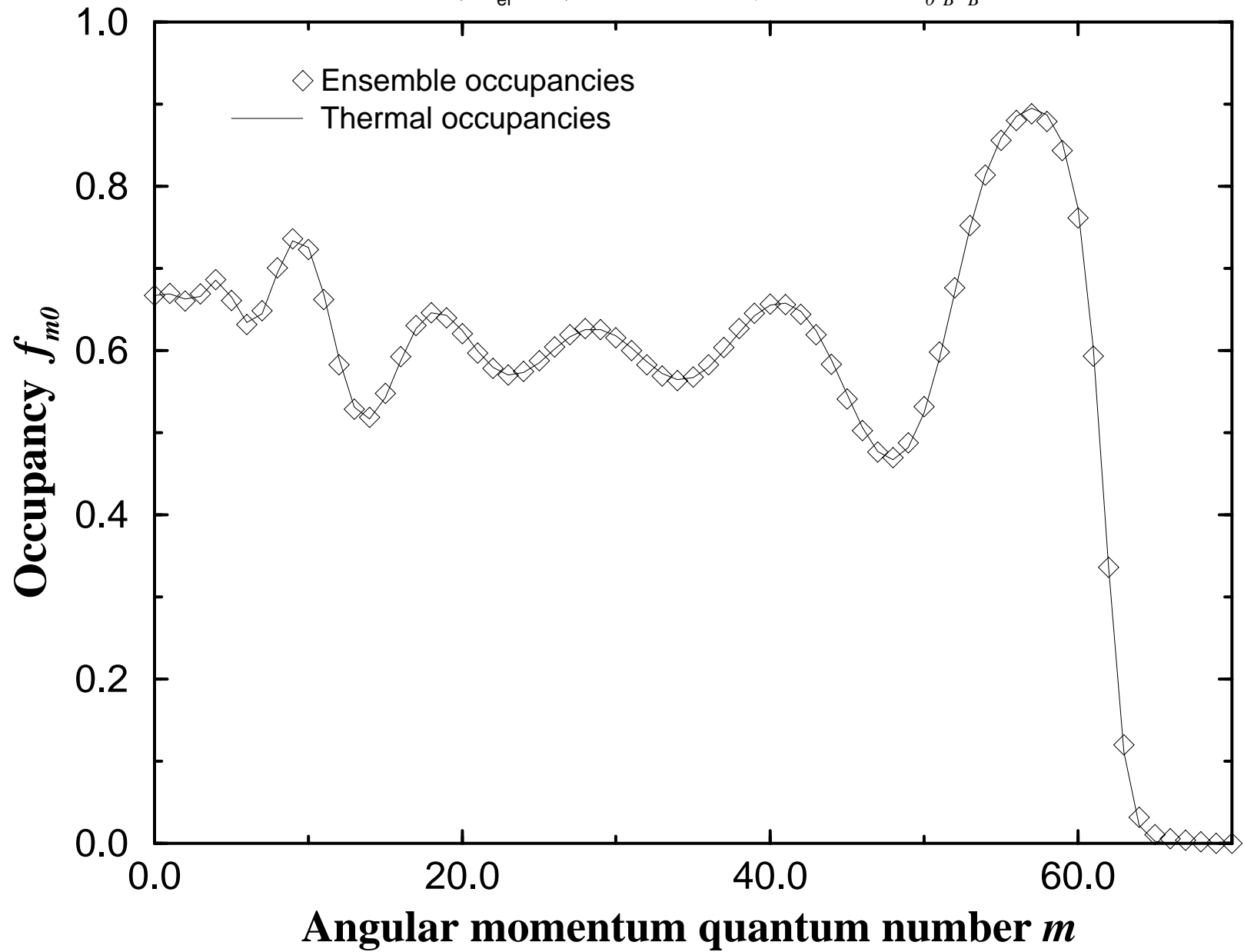
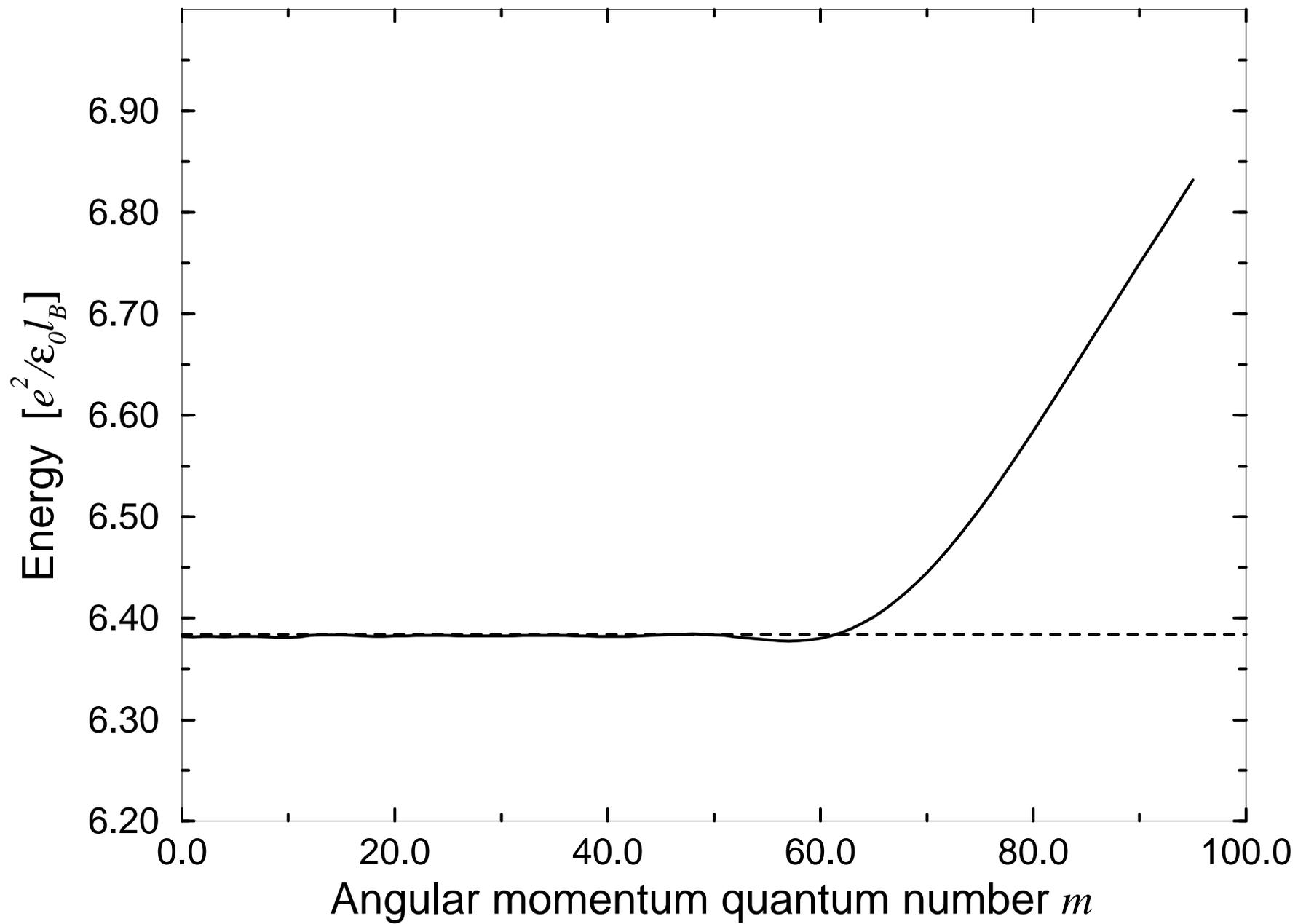
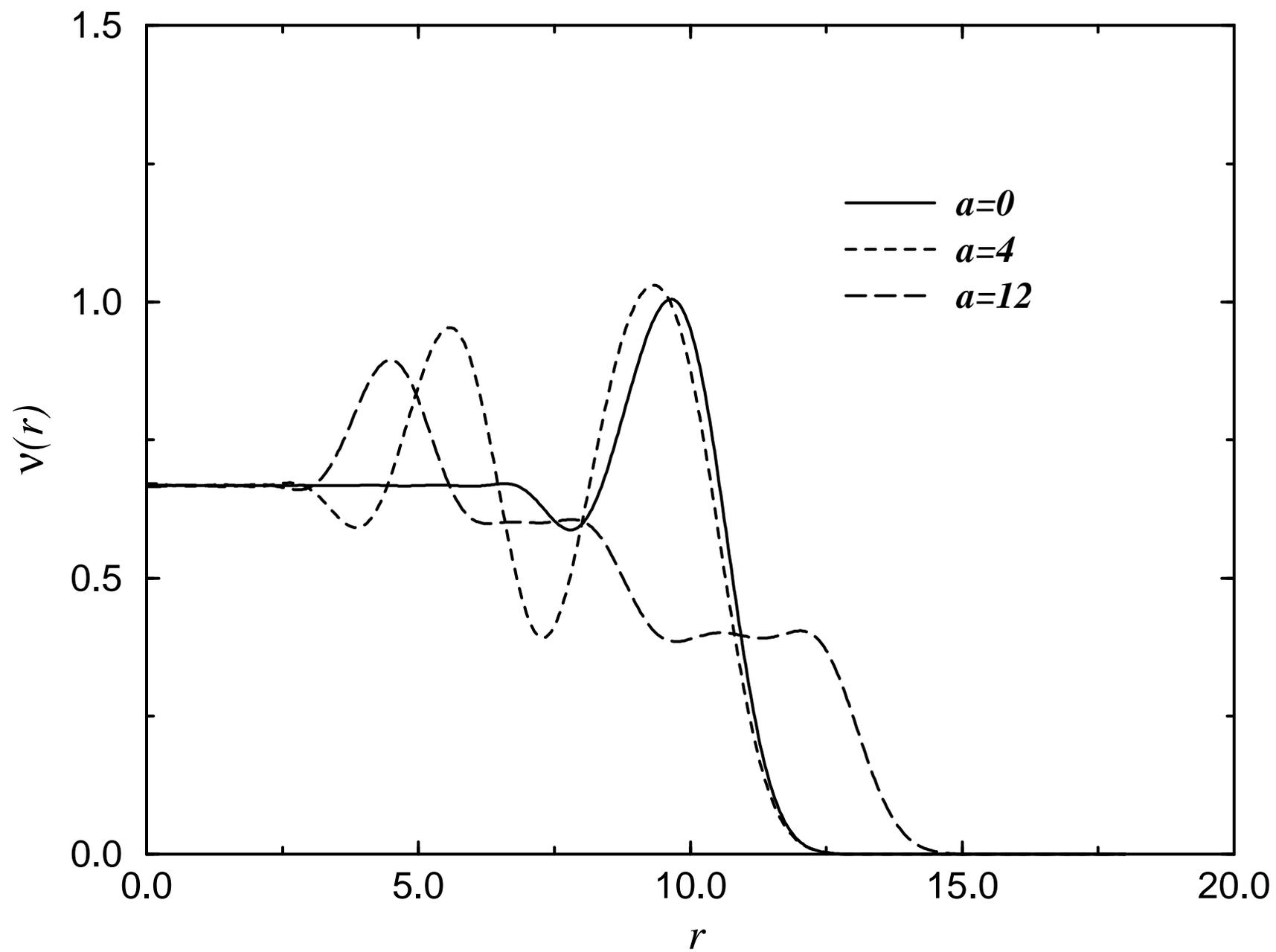
\end{document}